# Disarming the Ultimate Historical Challenge to Scientific Realism

Peter Vickers

### ABSTRACT

Probably the most dramatic historical challenge to scientific realism concerns Arnold Sommerfeld's ([1916]) derivation of the fine structure energy levels of hydrogen. Not only were his predictions good, he derived exactly the same formula that would later drop out of Dirac's 1928 treatment (something not possible using 1925 Schrödinger–Heisenberg quantum mechanics). And yet the most central elements of Sommerfeld's theory were not even approximately true: his derivation leans heavily on a classical approach to elliptical orbits, including the necessary adjustments to these orbits demanded by relativity. Even physicists call Sommerfeld's success a 'miracle', which rather makes a joke of the so-called 'no miracles argument'. However, this can all be turned around. Here I argue that the realist has a story to tell vis-à-vis the discontinuities between the old and the new theory, leading to a realist defence based on sufficient continuity of relevant structure.

## 1 Introduction

The historical challenge to scientific realism widely considered the most serious and problematic concerns Sommerfeld's ([1916]) derivation of the fine structure spectral lines of hydrogen.[1] What makes the case so powerful against realism is how it apparently overcomes the usual realist defences. For example, when it comes to the success-to-truth inference at the core of scientific realism, modern realists set a high bar for the level of success required. Novel predictive success has long been favoured over 'mere' explanatory success, but more recently it has become clear that even novel predictive success should not be motivating if the predictions are vague or otherwise unimpressive (despite their novelty). In addition most contemporary realists insist that such success only justifies a realist commitment to the 'success-fuelling' or 'working' parts of the theory in question; the wider theory might well be radically false, without any threat to realism. But these popular defences do not seem remotely helpful when it comes to the Sommerfeld case. Sommerfeld's 'fine structure formula' was perfect, since it is exactly the same formula that later resulted from the relativistic Dirac quantum mechanics (QM) of 1928 (itself an improvement on Schrödinger–Heisenberg QM). And at the very heart of Sommerfeld's theory are continuous worldline elliptical orbits of electrons, derived using relativistic classical mechanics. Sommerfeld assumed that the mass of the electron changes as its velocity changes during its orbit, in line with relativity. But as Griffiths ([2004], p. 16) notes in his popular textbook: 'It's not even clear what velocity *means* in QM'.

---

[1]  For example, at the conference The History of Science and Contemporary Scientific Realism, held in Indianapolis 19–21 February 2016, this case came up repeatedly during both formal and informal discussion, as a case apparently impossible to reconcile with the realist's success-to-truth inference.

Many figures in the contemporary realism debate do not like to talk in terms of counterexamples to realism. Instead realism is said to involve a defeasible commitment: scientific success is (highly) indicative of truth, but does not guarantee it. But this overlooks the fact that the measure of scientific success, including predictive success, is a matter of degree (cf. Fahrbach [2011]). The more impressive the success, the closer the realist comes to inferring that that success must be born of truth. Thus one can see why some have been tempted to refer to the Sommerfeld case as a 'counterexample', at least loosely speaking: the quantitative accuracy of Sommerfeld's fine structure formula is extremely impressive, and thus highly motivating for the realist. Thus it seems especially hard in this case for the realist to shrug her shoulders and say 'Well, this is just one case, and my inference is defeasible'

Despite the apparent measure of the success, certain anomalies provide a possible reason for the realist to evade making a commitment in this case. This is explored in Section 2, but I argue that, all things taken into consideration, the realist must make a commitment. Even if the realist could find a way to avoid making a commitment, there remains the mystery of why Sommerfeld's radically false theory led to the 'perfect' formula. In the physics literature the case is known as 'The Sommerfeld puzzle', and two physicists in particular have attempted to solve the puzzle, whilst several others have commented on it. The physics literature is of crucial significance for the philosophical debate, and I turn to it in Section 3. I then build up to an explanation of Sommerfeld's success, which also stands as a realist defence. In Section 4 I start by turning to the non-relativistic case, and the predictive success of old quantum theory (OQT) vis-à-vis the spectral lines of ionized helium. I offer a new approach to this case, which then stands as a platform from which to approach the relativistic case and the fine structure formula in Section 5. Section 6 tackles some outstanding questions concerning the structural relationship between the old and the new theory; Section 7 is the conclusion.

## 2 No Realist Commitment Required?

In (Sommerfeld [1916])—building on Bohr's 1913 model of the hydrogen atom—Sommerfeld derived the fine structure formula for the allowed energy states of unperturbed hydrogen, and thus via $\Delta E = h\nu$ the possible frequencies of the hydrogen spectral lines (for $Z = 1$)[2]:

$$E_{n_r, n_\varphi} = m_0 c^2 \left[ 1 + \frac{\alpha^2 Z^2}{[n_r + (n_\varphi^2 - \alpha^2 Z^2)^{1/2}]^2} \right]^{-1/2}. \quad (1)$$

*Prima facie* this result ticks all the boxes the realist requires to make a doxastic commitment. The formula (combined with $\Delta E = h\nu$) encodes countless novel predictions of spectral lines with extreme quantitative accuracy. It even applies to different elements (not only hydrogen), so long as the atoms are ionized such that they are one-electron atoms. In particular, the formula applies to the ionized helium fine structure ($Z = 2$).

However, the realist might offer two separate reasons that a realist commitment to (even parts of) Sommerfeld's ([1916]) theory was in fact not warranted: (i) the scientific landscape was changing rapidly at that time (1916–25), and perhaps a cautious realist should not make a realist commitment to any theory, however successful, until a few years have passed and the dust has settled (cf. Harker [2013], Footnote 24); (ii) although Sommerfeld's success vis-à-vis the fine structure formula was widely perceived as excellent, the theory also encountered some anomalies, and a realist inclined towards credence updating might insist that after iterating one's degree of belief in the theory given all available evidence, including anomalies/disconfirmations, one's final degree of belief might not be very high.

Concerning reason (i), it does seem reasonable that a cautious realist will hold back from making a commitment whenever the relevant science is in serious flux. There is of course the question of how long the realist should hold back. The appropriate length of time surely depends on the extent of the scientific turmoil, and the measure of the relevant success that is calling for a realist commitment. Here we meet with a new version of Stanford's ([2009], p. 384) 'threshold problem'. Stanford's worry was that realists can just keep raising the bar for what counts as sufficient success, so as to insulate realism from problematic historical challenges. The new version of this worry pertains to how long the realist waits, following significant success, before she makes a commitment.

---

[2] In this equation, $m_0$ is the rest mass of the electron, $c$ is the speed of light, $\alpha$ is the fine structure constant equal to $e^2/\hbar c$, $n_r$, and $n_\varphi$ are the radial and angular quantum numbers, and $Z$ is the proton number. By 'unperturbed' here and elsewhere I mean 'not affected by electric or magnetic fields'.

To dodge the Sommerfeld challenge in this way, the realist would need to insist that it was prudent to wait at least ten years following Sommerfeld's ([1916]) success before making a realist commitment. But can that be justified by the measure of the scientific flux during that period? Here a sharp distinction needs to be made between Sommerfeld's theory as a theory of one-electron atoms (especially those unperturbed by magnetic fields), and Sommerfeld's theory as a theory of atoms generally. Construed as a theory of atoms generally, there was indeed significant theoretical flux between 1916 and 1925: the wider research programme known as OQT struggled with relevant phenomena for neutral helium and all heavier elements, and this led to all sorts of different theoretical suggestions in an effort to achieve empirical adequacy.[3] And even when we focus only on one-electron atoms (especially hydrogen), there were problems accounting for magnetic field effects on the spectral lines (the Zeeman effects, including the Paschen–Back effect[4]). But construed as a theory of one-electron atoms unperturbed by magnetic fields, the theoretical flux was much more limited. And for current purposes the relevant conceptualization of 'the theory' is the narrow one that ignores heavier elements and magnetic fields, because Sommerfeld's ([1916]) fine structure success related specifically to unperturbed one-electron atoms.

At this point reasons (i) and (ii), above, cannot be separated: the measure of theoretical flux goes hand in hand with the measure of empirical adequacy (weighing up successes and anomalies). The realist surely has a good argument (on both counts) when it comes to OQT as a theory of atoms generally: there was a significant lack of empirical adequacy and corresponding theoretical flux as scientists attempted to achieve empirical adequacy. But construed as a theory of one-electron atoms unperturbed by magnetic fields the realist's case is much weaker. The measure of the empirical success vis-à-vis the hydrogen atom meant that there was relative theoretical stability. Anomalies cropping up for heavier elements could easily be put down to additional assumptions required to handle such elements, especially assumptions concerning how the multiple electrons in the atoms of heavier elements interact with each other.

---

[3] See, for example, (Mehra and Rechenberg [1982], Part 2).
[4] On the status of the Paschen–Back effect during the relevant period, see, for example, (Kragh [1985], pp. 102–6). By contrast the Stark effect was considered a great success of OQT (see, for example, Duncan and Janssen [2014]).

The realist can reply by pointing out that there were in fact some anomalies—with corresponding theoretical flux—even concerning Sommerfeld's theory of the unperturbed hydrogen atom. On this point the literature is actually quite misleading, often stating (or suggesting) that Sommerfeld's theory was 'in full agreement with observation' (Jammer [1966], p. 92).[5] Certainly Sommerfeld derived the 'perfect' fine structure formula (Equation (1)), which was later to emerge from 1928 Dirac QM, but this formula only gives the allowed energies of the hydrogen atom, and doesn't by itself deliver empirical results. The allowed energies only transfer into testable predictions once one adds two further ingredients: (i) the formula connecting energy differences with spectral line frequencies, $\Delta E = h\nu$, and (ii) a statement concerning which energy transitions (electron jumps) will actually occur. Now, there was no flux concerning (i), but (ii) was a matter of significant controversy during the period 1916–26. If the default assumption is that all such transitions will occur—what Kragh ([1985], p. 79) calls the 'primitive theory'—then the predictions definitely are not empirically adequate (Kragh [1985], p. 71).

In an effort to achieve empirical adequacy, scientists introduced 'intensity rules' for the expected intensity of the lines that do occur, with 'selection rules' as special cases where sometimes the intensity is zero because the transition in question never occurs. Kragh ([1985]) identifies six different versions of Sommerfeld's theory that were considered between 1916 and 1926, all differing only according to the intensity/selection rules in play.[6] Every version of Sommerfeld's theory gives different empirical results for the spectral lines and each was imperfect in one way or another. Thus we do in fact find theoretical flux on the precise issue in question during the precise period in question, and in addition the success of the theory is something less than is often claimed. During the period in question the match between theory and experiment—even concerning hydrogen and other one-electron atoms—could never be called 'perfect', and in practice anomalies were either ignored (Kragh [1985], pp. 80, 103) or explained away (p. 105).

The anomalies in question no doubt cast a question mark over the theory. But then again, a doxastic commitment is often thought to be warranted in the face of ('normal') anomalies so long as the successes are good enough. And mainstream scientists of the day did think the successes were good enough. Kragh ([1985]) writes that mainstream physicists 'were completely satisfied with Sommerfeld's theory of the hydrogen spectrum' (p. 102), concluding that 'Sommerfeld, Bohr, and their disciples decided that Paschen's confirmation of the theory was so decisive that no counter-evidence could qualify as serious anomalies' (p. 84). Thus, 'Despite the anomalies that turned up in the theory of the hydrogen spectrum, mainstream physicists continued to believe in the complete truth of Sommerfeld's explanation' (p. 117).[7] Even in February 1925, despite all the problems that were emerging for the OQT, Heisenberg wrote 'the hydrogen atom is in good shape' (p. 117).

---

[5] Cf. Keppeler ([2003a], p. 42), 'The success was overwhelming', and Granovski ([2004], p. 524), '[Sommerfeld's theory] somehow turned out to be equivalent to the consistent Dirac theory'.

[6] Five of these theories are listed in (Kragh [1985], p. 74); the sixth is the 'primitive theory'.

If we take seriously the idea that every scientific theory ever put forward will have some anomalies—that anomalies are indeed 'normal'—then the realist cannot seriously insist that a realist commitment will only be warranted when there are no anomalies. So can the realist argue that the anomalies here were significant enough to undermine the success? This is a very shaky line of defence for the realist given that there was a consensus amongst mainstream physicists of the day that the successes of the theory (construed narrowly as a theory of unperturbed one-electron atoms) totally overwhelmed the anomalies. The modern realist cannot really immerse herself in the history to such an extent that her view on the relationship between theory and evidence should be preferred over the view of the scientific community living through that period. Especially since the realist clearly has an agenda, and of course the realist knows that the theory ultimately failed, both of which are very likely to bias one's perspective.

Thus I suggest that it isn't reasonable for the realist to claim that no realist commitment was demanded by Sommerfeld's ([1916]) derivation of the fine structure formula. Yes, there were anomalies, but not of sufficient significance to worry mainstream physicists. And yes, there was some theoretical flux—even for the theory of the unperturbed hydrogen atom—concerning the 'selection rules' for energy transitions. But there was also very significant theoretical stability (when it comes to the hydrogen atom) between 1916 and 1925.

Thus the realist needs to accept that a realist commitment is warranted by this case, and turn her attention to what kind of realist commitment (if any) is compatible with the radical shift in thinking that occurred between 1916 (Sommerfeld) and 1928 (Dirac). In this project she doesn't need to start from scratch: the physics community has long been interested in the question of how to explain Sommerfeld's success.[8]

---

[7] For a striking example of this attitude from one such disciple considering the Paschen–Back anomaly in 1924, see (Kragh [1985], p. 106): 'Once again Paschen's measurements had made Sommerfeld's theory immune to criticism'

[8] What of philosophical literature on this puzzle? Here there is very little of any significance or substance. One paper that looks initially promising is (Hettema and Kuipers [1995]). But on closer inspection this article is asking a very different question, concerning the relationships between the theories of Rutherford, Bohr, and Sommerfeld. Only in a footnote at the very end of their conclusion do the authors speculate concerning the question of 'how to compare the 'old' and the 'new' quantum theory' ([1995], p. 295, Footnote 13).

## 3 Enter the Physicists

In the physics literature this case is known as 'The Sommerfeld puzzle'. By far the most common opinion coming from the physicists is that Sommerfeld's success was a 'fluke', a 'coincidence', and a 'lucky accident'. For example, Kronig ([1960], p. 9) describes it as 'perhaps the most remarkable numerical coincidence in the history of physics'. This take on the puzzle suggests the realist shouldn't hope to find an explanation in terms of the truth content of Sommerfeld's theory; instead, we are encouraged to accept that Sommerfeld was just (incredibly) lucky. Curiously, physicists have often described this case as a 'miracle', and even as a 'cosmic joke', directly contradicting (unintentionally!) the 'no miracles' or 'no cosmic coincidences' argument for scientific realism.[9]

However, dismissing it as a lucky accident doesn't really explain how Sommerfeld's ([1916]) assumptions could lead to the same predictions. Several physicists have tried to provide more by way of explanation. Most popular here is a 'two errors cancelling out' explanation. Yourgrau and Mandelstam ([1968])—possibly influenced by Schiller ([1962], p. 1108)—have been influential, concluding their discussion with, 'Sommerfeld's explanation was successful because the neglect of wave mechanics and the neglect of spin by chance cancel each other in the case of the hydrogen atom' ([1968], p. 115). Other physicists expressing similar thoughts include Eisberg and Resnick ([1985], p. 286) and Keppeler ([2004], p. 49). And indeed, one might think this is the only possible explanation. Certainly Sommerfeld was working with classical mechanics, as opposed to the wave mechanics of the modern quantum theory. And in addition it is clear that Sommerfeld did not include anything like 'spin' in his theory. If two crucial ingredients were missing, but the exact correct result was nevertheless achieved, then isn't it clear that these two things exactly cancel each other, at least when it comes to the hydrogen atom?

Another option is that these two things are completely idle when it comes to the fine structure. This would be much better for a realist explanation, since she could then hope to explain Sommerfeld's success by noting that he left out things that are irrelevant vis-à-vis the final result, and this is perhaps consistent with the conclusion that Sommerfeld's theory includes sufficient truth to reach the correct predictions. However, this seems hopeless when one notes that everywhere in the physics literature the very cause of the fine structure is said to be the spin; to give just one example: 'The interaction between the spin and the electron's orbit is called spin-orbit interaction, which contributes energy and causes the fine-structure splitting' (Letokhov and Johansson [2009], p. 37). But if the spin is the cause, then it definitely makes a difference, and is not idle. Thus we seem surely led to the conclusion that the only way Sommerfeld could have been successful is if his neglect of spin and his use of the wrong mechanics cancel each other out exactly. Such a 'freak of nature' coincidence (Keppeler [2003a], p. 68) certainly does not seem conducive to scientific realism!

---

[9] For some nice quotes to this effect (in addition to Kronig), see (Rozental [1967], p. 73, quotation from the correspondence of Carl Oseen; Heisenberg [1968], p. 534; Biedenharn [1983], p. 14, also including a 1956 quotation from Schrödinger; Eisberg and Resnick [1985], p. 286; Brown *et al.* [1995], p. 92; Keppeler [2003a], pp. 68ff, [2003b], p. 86; Granovski [2004], p. 524). See also (Kragh [1985]; Eckert [2013], pp. 426ff). Sommerfeld himself apparently didn't see a 'lucky coincidence' here, instead emphasizing (in 1940 and in 1942) that both theories make essential use of special relativity (see Kragh [1985], pp. 124ff; Eckert [2013], p. 427). But as Eckert ([2013]) rightly notes, the common use of special relativity in the two theories 'was no explanation'.

Some hope for the realist remains here, however, since the above discussion brushes over a couple of things. Talk of 'two errors cancelling' suggests that the errors are independent of each other, with one contributing a certain quantifiable error and the other taking that error away. That is misleading, since in Dirac QM the spin is intimately related to the relativistic wave mechanics, and does not have to be introduced as a separate assumption at all: one can't possibly consider what difference the spin makes to Dirac QM, since one can't do Dirac QM without the spin operator.[10] In addition, a structural realist might wonder whether there is some deep structural correspondence between the two theories, and this might be established in a purely formal way, with no discussion of 'spin' required. Certainly some physicists have been inclined in this direction. Heisenberg was curious about this puzzle, and wrote 'It would be a stimulating project to explore whether this is truly a miracle, or whether perhaps the group-theoretical structure of the problem underlying the formulations of both Sommerfeld and Dirac itself leads already to this formula' ([1968], p. 534). Unfortunately, Heisenberg never took up this 'stimulating project'. But in 1982–3 Lawrence Biedenharn did.

Biedenharn ([1983], p. 14) at first appears to do everything the structural realist would wish. He claims to 'resolve' the puzzle, and he does this in terms of an 'underlying symmetry of the problem'. He asserts that 'Sommerfeld's success is not at all a matter of blind luck' (p. 14), and indeed claims to demonstrate that 'the argument of Mandelstam and Yourgrau (that Sommerfeld succeeded 'because the neglect of wave mechanics and the neglect of spin effects by chance cancel each other') cannot possibly be correct' (p. 30). He reaches this conclusion by arguing that 'wave mechanics *per se* makes *no change* in the answer' (p. 30). And if the wave mechanics is idle vis-à-vis the final predictions, then it can't introduce an error that cancels out the error due to neglecting the spin.

---

[10] The Dirac derivation begins with the Hamiltonian $H = \rho_1 \sigma \cdot pc + \rho_3 m_0 c^2 - \frac{Ze^2}{r}$, where $\sigma$ is the spin operator, and $\rho_1$ and $\rho_3$ are $4 \times 4$ matrices. One then solves the eigenvalue problem $H\psi = E\psi$, where $\psi$ is a four-component spinor; see, for example, (Biedenharn [1983], pp. 25ff).

*Disarming the Ultimate Historical Challenge* 9However, when we look harder at (Biedenharn [1983]) things appear to go wrong for the realist. First of all, Biedenharn's argument that wave mechanics *per se* makes no change to the answer is based on comparing non-relativistic OQT with non-relativistic Schrödinger–Heisenberg QM. That is to say, Biedenharn shows that one gets the same Bohr energies for the allowed states of the electron whether you make use of classical mechanics or whether you make use of wave mechanics. But does this result obviously carry across to the relativistic case, comparing Sommerfeld with Dirac? The comparison simply can't be done, since (as noted above) in Dirac QM spin and relativity are intimately intertwined, such that it is impossible to consider Dirac QM without spin and check to see whether the fine structure formula still results. And one might have principled reasons for doubting that the non-relativistic result carries across to the relativistic case. In particular, the underlying $O(4)$ structural symmetry of the hydrogen problem is lost when one turns to the relativistic case.

There is also a basic anti-realist objection to Biedenharn's analysis, construed as a defence of selective realism. The objection is simply that for selective realism to work here we'd need to see appropriate continuity of 'working parts' despite the radical differences between the two theories. Since continuous worldline elliptical orbits are not even approximately involved in Dirac QM, and since spin is not even approximately involved in Sommerfeld's theory, the selective realist needs to show two things:

(1) that Sommerfeld's derivation can succeed without any explicit mention of, and without implicit reliance upon, electron orbits, and,

(2) that spin is not essential to the modern derivation (or that it is somehow hiding in Sommerfeld's ([1916]) assumptions).

But in Biedenharn's analysis there appears to be no attempt to do either of these things: his Sommerfeld derivation is full of references to orbits, and his Dirac derivation is full of references to spin. He even states at one point 'The spin is of course an essential ingredient in the relativistic quantal calculation' ([1983], p. 15).

One may wonder at this point whether, in his capacity as a theoretical physicist, Biedenharn just has a completely different agenda to that of the philosopher. Certain passages in his paper suggest an agenda very closely related (at least) to that of the scientific realist, but in the end there are difficulties that seem to close the door on a realist who would wish to simply present the paper as a selective realist defence. And the realist philosopher has quickly run out of physicist allies, since no other physicist has tried to 'resolve' the puzzle in the sort of way Biedenharn does.

With the realist on the ropes the anti-realist might now introduce what would appear to be the killer blow. Keppeler ([2003a], [2003b], [2004]) does not merely advocate a coincidental 'cancelling out' explanation, as some physicists do; he apparently demonstrates it. In (Keppeler [2003a], [2003b]), he shows that Sommerfeld's quantum conditions are missing a ½ 'Maslov' term, and he compares this with the influence of spin. After some lengthy calculations, he concludes 'in this particular situation the contribution deriving from the Maslov index and the influence of spin accidentally cancel each other' (Keppeler [2003b], p. 86). Thus it seems a 'coincidental cancelling out' explanation must be accepted, and it's the end of the road for a realist explanation. Or is it?

## 4 A New Approach to the Non-relativistic Success

My realist defence begins with a new approach to explaining the non-relativistic success of OQT that emerged in the years 1913–15. Other papers (Vickers [2012]; Ghins [2014]) have offered selective realist defences specifically against Bohr's 1913 (non-relativistic) model of the atom construed as an example of significant empirical success issuing from (radically) false hypotheses. But these defences are unhelpful when it comes to the Sommerfeld challenge. In this section, I briefly explain why they are unhelpful, before moving on to consider a new theoretical approach to the hydrogen atom that had emerged by 1915, and which presents a new (non-relativistic) challenge to the realist. I then provide a realist defence against the 1915 challenge that, by its nature, also stands as a clear springboard for tackling Sommerfeld's ([1916]) relativistic extension, and the fine structure success.

In my (Vickers [2012]), I consider both Bohr's 1913 success and Sommerfeld's 1916 success in the context of the scientific realism debate. The take-home message is that there is a way to explain Bohr's success in terms of truth using the selective realist strategy, but that 'this strategy almost certainly can't work for Sommerfeld's derivation' (Vickers [2012], p. 3). One may ask why it can't work for Sommerfeld's derivation, given that Sommerfeld's theory is just a de-idealization of Bohr's theory. There is a clear answer to this question: the application of the selective realist strategy to Bohr's successful prediction of the spectral lines of ionized helium found in (Vickers [2012]) makes essential use of the Balmer formula. This is a purely phenomenological formula for the 'gross structure' spectral lines of hydrogen, which Bohr made use of in a lecture at the end of 1913 in order to derive a theoretical formula for the Rydberg constant. This can then be used to predict the 'gross structure' spectral lines of ionized helium. Drawing on (Norton [2000]), I argued (Vickers [2012]) that the Balmer formula, combined with a subset of Bohr's theoretical commitments that are all true (or very approximately true) in light of modern QM, leads to his true predictions ([2012], pp. 8ff). However, there is absolutely nothing like the Balmer formula for the fine structure spectral lines. So there is no chance of making the same sort of argument for Sommerfeld's derivation of those lines.

At first this seems like yet another reason to suppose that a realist explanation of Sommerfeld's success will not be forthcoming: the selective realist strategy, as it has been applied to Bohr's 1913 success, can't possibly work for the Sommerfeld success. But could there be another, different realist approach to the non-relativistic success, which perhaps could carry across quite naturally to the Sommerfeld success? Well, there actually needs to be such a different realist approach to the non-relativistic success, because there are other derivations I did not consider in my (Vickers [2012]). In a footnote ([2012], Footnote 13), I do acknowledge the fact that I focus on the third derivation Bohr offered in 1913. Now, that is reasonable as far as it goes, since the first two derivations were not particularly impressive or influential, and the realist can and should ignore them.[11] But by 1915 it became possible to derive the ionized helium spectral lines from OQT in a quite different way, without assuming the Balmer formula, and apparently putting more weight on electron orbits. Here we meet with a new problem for the realist, since it turns out that (Vickers [2012]) has done only half a job, and the realist needs to revisit the non-relativistic success. However, a tiny glimmer of hope for the realist also opens up at this point: if there can be a realist explanation for the success of the 1915 derivation, then just possibly this same explanation could carry across to Sommerfeld's ([1916]) relativistic extension of the theory, and the fine structure success.

A crucial step in developing a derivation of Bohr's successes that does not depend on assuming the Balmer formula (where the Balmer formula is instead derived) was to generalize Bohr's quantum condition in such a way that the ideas of Bohr and Planck could be unified. This came in 1915, courtesy of Sommerfeld, Wilson, and Ishiwara (see Heilbron and Kuhn [1969], p. 280). The resultant 'phase-integral' quantum conditions—the so-called BWS conditions (after Bohr, Wilson, and Sommerfeld)—were not developed especially to apply to circular orbits. Instead, by 1915 it was usual to employ them in a derivation that assumed elliptical orbits and thus two degrees of freedom (see, for example, Sommerfeld [2014]),

$$\oint p_\varphi d\varphi = n_\varphi h \quad (2)$$

$$\oint p_r dr = n_r h \quad (3)$$

thus introducing quantum numbers $n_\varphi$ and $n_r$.[12] With these conditions in place the stage was set. A convenient early reconstruction of the 1915 derivation of the

---

[11] As Arabatzis ([2006]) notes, the first two derivations of 1913 are 'imperfect' (p. 141) because 'based on questionable foundations' (p. 130), something that Bohr himself acknowledged. See (Heilbron and Kuhn [1969], pp. 266ff, especially p. 270) for full details.

[12] In the special case where the orbital radius is constant, Equation (2) is equivalent to Bohr's original quantization of angular momentum. Equation (3) was sometimes called 'quantization of ellipses' or 'quantization of eccentricity'.

Bohr energies is given in (Pauling and Goudsmit [1930], pp. 13–20), which I now follow quite closely using the abbreviation 'PG'.

The derivation of the Bohr energies (with energy $E$ and Rydberg constant $R$),

$$E_n = -\frac{RhcZ^2}{n^2} = -\frac{2\pi^2 m_0 Z^2 e^4}{n^2 h^2}, \qquad (4)$$

has two central physical assumptions: (i) the Coulomb Hamiltonian of classical mechanics, and (ii) the BWS quantum conditions. We start with the Coulomb Hamiltonian, expressing the potential and kinetic energy in radial coordinates $r$ and $\varphi$ ([1930], p. 14). We then make some manipulations to express the Hamiltonian in terms of the momenta $p_r$ and $p_\varphi$ associated with the two degrees of freedom $r$ and $\varphi$. The angular momentum $p_\varphi$ is seen to be a constant, so we just write it as $p$, and the radial momentum $p_r$ can be expressed in terms of $p$. With some further manipulations we can derive the following ([1930], p. 15),

$$\left(\frac{1}{r}\frac{dr}{d\varphi}\right)^2 = \left(\frac{2m_0 Z e^2}{p^2}\right)r + \left(\frac{2m_0 E}{p^2}\right)r^2 - 1, \qquad (5)$$

which takes exactly the same form as the equation for an ellipse that drops out of pure geometry (with $a$ as the major axis and $\varepsilon$ the eccentricity):

$$\left(\frac{1}{r}\frac{dr}{d\varphi}\right)^2 = \left(\frac{2}{a(1-\varepsilon^2)}\right)r - \left(\frac{1}{a^2(1-\varepsilon^2)}\right)r^2 - 1. \qquad (6)$$

A comparison of the terms on the RHS of these two Equations (5) and (6) gives us the following equalities (p. 16):

$$a(1-\varepsilon^2) = \frac{p^2}{m_0 Z e^2}, \qquad (7)$$

$$\frac{2m_0 E}{p^2} = -\frac{1}{a^2(1-\varepsilon^2)}. \qquad (8)$$

Now some cancelling gives us our first equation for the energy of a given orbit:

$$E = -\frac{Ze^2}{2a}. \qquad (9)$$

This expression for $E$ is already of interest, since we see that the energy depends only on the major axis of the ellipse $a$. It does not depend on the eccentricity, thus helping to explain why Bohr's work in terms of circular orbits was so successful (he ignored something that happened to be irrelevant).

We now bring in the BWS quantum conditions—Equations (2) and (3)—thus quantizing the angular and radial momenta (p. 17). Since the angular momentum is a constant we reach:

$$p = \hbar n_\varphi, \tag{10}$$

where $n_\varphi$ is our first quantum number. When it comes to quantizing the radial momentum things are trickier, but after a few manipulations (using the relation between $p$ and $p_r$) we reach

$$1 - \varepsilon^2 = \frac{n_\varphi^2}{(n_r + n_\varphi)^2}, \tag{11}$$

where our first quantum number $n_\varphi$ features again because we introduced $p$ during the manipulations (not shown) and we know that $p = \hbar n_\varphi$.

We can now go back to Equations (7) and (8), this time equipped with Equations (10) and (11) which give us handy substitutions for $p$ and $(1 - \varepsilon^2)$. Some substitutions, manipulations, and eliminations result in the Bohr energies (Equation (4)). The identity depends on introducing a new quantum number $n = n_r + n_\varphi$, which just helps to simplify things, showing clearly that the allowed energies only depend on one quantum number. And of course, plugging $Z = 2$ into Equation (4), and using $\Delta E = h\nu$, quickly delivers the ionized helium spectral lines. Thus we have a route to Bohr's most impressive predictive successes that makes essential use of classical mechanics and does not depend on assuming the Balmer formula.

Thus robbed of the explanation in (Vickers [2012]), how can the realist explain the success of this derivation in terms of truth? A selective realist inference from the success to the approximate truth of the working parts would suggest a (potentially problematic) realist commitment to both classical mechanics and the BWS quantum conditions. These are the two central physical assumptions going into the derivation, and indeed they are *prima facie* doing essential work to fuel the derivation. But can the realist reasonably claim that they are approximately true? Or at least appropriately linked with the truth via a structural relationship?

Here begins my argument for a realist explanation of Sommerfeld's fine structure success. It starts by noting a structural relationship between non-relativistic classical mechanics (as used in the PG derivation) and the non-relativistic Schrödinger–Heisenberg QM that emerged in 1925. In particular, it starts by noting that there is some excess structure in QM that is completely idle when it comes to deriving the Bohr energies. And it just so happens that when one removes this excess structure one recovers something almost identical to one formulation of (non-relativistic) classical mechanics.

No doubt there are fundamental differences between the physics of classical mechanics and the physics of QM. But if the scientific realist may be permitted to focus the structural relationship between the two, things look very good for her, since there is a crucial sense in which the relationship is extremely intimate. As shown in (Sakurai [1985], pp. 103ff), for example, the time-dependent Schrödinger equation reduces to the Hamilton–Jacobi formulation of classical mechanics in the semi-classical limit $\hbar \to 0$. Goldstein ([1980], p. 491) shows this clearly by presenting the following version of the Schrödinger equation,

$$\frac{1}{2m}(\nabla S)^2 + U + \frac{\partial S}{\partial t} = \frac{i\hbar}{2m}\nabla^2 S, \qquad (12)$$

which reduces to the Hamilton–Jacobi equation if the RHS equals zero, interpreting $S$ as Hamilton's principal function. And the RHS does equal zero when we take the semi-classical limit $\hbar \to 0$.

Of course semi-classical QM and classical mechanics, despite sharing an equation, are not the same thing. In particular, in the Hamilton–Jacobi formulation of classical mechanics, although the motion of a particle is represented mathematically as a wave, the wave mechanics involved is to be interpreted purely as a representational device, without physical significance. In semi-classical QM we have a different perspective, which invites boundary conditions that dictate how the quantum numbers are introduced. This leads to quantum conditions that differ from the BWS conditions (Equations (2) and (3)), since they include a ½ 'Maslov' term as follows[13]:

$$\oint p_\varphi d\varphi = (n_\varphi + 1/2)h, \qquad (13)$$

$$\oint p_r dr = (n_r + 1/2)h. \qquad (14)$$

Now we ask the question whether one recovers the same Bohr energies (Equation (4)) with a semi-classical approach, and the answer is 'yes' (see, for example, Schiller [1962], pp. 1105–6).

Is this good news for the scientific realist? Not yet, since there are different possible interpretations of what we've just seen. One option is a 'coincidental cancelling out' explanation even for this non-relativistic success: the PG derivation makes use of the wrong mechanics, and the wrong quantum conditions, but it delivers the correct result anyway. So quite naturally one might think that since there are two mistakes, but the end result is not affected, there must be a lucky cancelling out of the errors. If this were right the success of the 1915 derivation would have to be put down to luck, not to truth.

---

[13] Useful sources are (Sakurai [1985], p. 107; Keppeler [2003b], p. 105; Müller-Kirsten [2012], p. 297, Equation 14.63b).

We can test this 'cancelling out' explanation directly. If there is a cancelling out of two errors, then each error can be considered separately, and we should be able to derive the actual term that is introduced by one error and cancelled out by the other. When it comes to the quantum conditions, we can re-run the PG derivation given above, but use the quantum conditions with the Maslov term (Equations (13) and (14)) instead of the BWS quantum conditions (Equations (2) and (3)). But in fact doing this makes no difference whatsoever to the final result! No error term affects the final result when we use the alternative quantum conditions in the PG derivation. All we have to do is make a different substitution $n = n_r + n_\varphi + 1$ in the final stages of the derivation (instead of $n = n_r + n_\varphi$), exactly as seen in (Schiller [1962], p. 1106). This doesn't change the allowed energies; it just tidies things up.

Now if the mistake with the quantum conditions is a mistake that doesn't make a difference vis-à-vis the resultant energies, then it can hardly play a role in cancelling out some other mistake. Thus the use of classical mechanics, as opposed to QM, can't make any difference either. And that is confirmed by the fact that, as seen above, when one takes the $\hbar \to 0$ limit of QM one (a) recovers the Hamilton–Jacobi equation, and (b) ends up with exactly the same Bohr energies; the term on the RHS of Equation (12) is idle structure vis-à-vis deriving the Bohr energies.

So we avoid a 'cancelling out' explanation of the non-relativistic success. But are we left with a realist explanation? The explanation we have reached includes two factors: (i) the PG derivation misses out some structure (the RHS of Equation (12)) that is idle vis-à-vis the result, and (ii) the PG derivation makes use of a quantum condition that is just one option for reaching the result. On (i), many realists will be satisfied, since the $\hbar \to 0$ relationship between classical and QM is reminiscent of many predecessor/successor theory relationships in the history of physics. The careful realist would never be so naïve as to suppose that, in a case of predictive success, there couldn't possibly be some further unidentified structure that just doesn't make a difference for the prediction(s) in question. Such caution from the realist is especially warranted when there are phenomena closely related to those successfully predicted that can't be similarly accommodated by the theory. And this was exactly the situation during the years of the OQT (spectral line intensities and the neutral helium problem, for example).

Turning to (ii), the realist can treat this in the same way, describing the ½ Maslov term in the semi-classical quantum conditions as 'idle structure': a theoretical difference that doesn't make an empirical difference vis-à-vis the Bohr energies. The realist might also point out that when it comes to the ½ Maslov term it was really quite easy to notice, in the context of the 1915 derivation, that such a term wouldn't make a difference to the final spectral line predictions. And whenever the realist recognizes that, at some step in a derivation, two different assumptions lead to the same conclusion, she has a principled reason to remain somewhat agnostic between those two assumptions (cf. Psillos [1999], p. 110). Of course, Sommerfeld and others had prior reasons for preferring the BWS quantum conditions (see Jammer [1966], pp. 90ff). But those other reasons must be divorced from the confirmation accrued to the conditions by the success. The confirmation accrued to the conditions by the success cannot be overly significant when there is another formula available that would work just as well. In the context of the realism debate, that means that one's degree of belief in those conditions should not increase dramatically as a direct result of the predictive success. Instead, the realist might sensibly opt to stay rather neutral between the two options.

With the 'cancelling out' explanation left behind, and with due care taken over the nature of the realist commitment warranted by the success, the realist has a story to tell vis-à-vis the derivation of the ionized helium spectral lines by use of classical mechanics and the BWS quantum conditions. So far this has all concerned the non-relativistic success, but the time is well spent. The stage is set to tackle the Sommerfeld miracle.

## 5 Relativity and Spin

I've argued that the realist has a good story to tell concerning the success of the 1915 derivation, a story that depends on appropriately relating the PG assumptions to the assumptions at the heart of non-relativistic QM. If this is accepted, the realist has a natural way to extend the story to include the fine structure success. In both the old and the new theory the fine structure formula follows if we make the necessary relativistic adjustments to the assumptions in play. The underlying nature of the structural relationship relied upon by the realist in the non-relativistic case will be preserved, since the structural adjustments required by relativity will affect both the old and the new theory in the same fundamental sense.

This is the short story. We can confirm that it is a good story in a number of different ways. For one thing, we might wonder about the identification of 'idle structure' in non-relativistic QM that crucially featured in the story told in the previous section. Can we identify 'idle structure' in exactly the same sort of way in the relativistic setting? With Dirac QM the theory has changed quite significantly, since now the relevant equation involves a four-component spinor, $4 \times 4$ matrices, and a spin operator (see Footnote 10). If the structural relationship between classical and QM noted in the previous section really is preserved in the relativistic setting, then we should expect there to be idle structure also within Dirac QM, and we should expect a semi-classical treatment of Dirac QM to still deliver the fine structure formula. And indeed, this is exactly what we do find; Keppeler ([2003a]) demonstrates that the fine structure formula (Equation (1)) still exactly follows when we take away from Dirac QM certain mathematical structure via the action of taking the semi-classical limit $\hbar \rightarrow 0$. Thus—when it comes to the specific issue of deriving the fine structure formula—there is excess structure in Dirac QM that is fully analogous to the excess structure we identified in the non-relativistic case.

But what about the role of spin in Dirac QM, and the fact that the fine structure is said to be caused by the spin-orbit interaction? If spin is an essential part of Dirac QM, how can the realist hope to find sufficient continuity through theory change to explain Sommerfeld's success? Two things need to be noted here: (i) spin is not assumed in Dirac QM—it is not one of the physical assumptions one makes within Dirac QM in order to reach the fine structure formula, and (ii) the realist must not commit to the physical reality of 'spin' anyway, regardless of the Sommerfeld puzzle.

To expand on point (i): the realist's success-to-truth inference has always been an inference from empirical success to the (approximate) truth of the assumptions essentially employed to generate that success.[14] In the case of Dirac's derivation of the fine structure formula (Equation (1)), it is simply not the case that one must essentially assume the existence of spin in order to generate the result. And since Dirac did not assume the existence of spin, it immediately makes more sense why Sommerfeld managed to derive the fine structure formula without assuming the existence of anything like spin. Instead, in both theories, the fine structure formula follows from a relativistic adjustment of the mechanics. It's just the case that in the Dirac theory the relativistic adjustment also leads to the introduction of a spin operator, which in turn suggests the reality of spin as a property of the electron.[15]

But how strongly does it 'suggest' the reality of spin? This question leads directly to point (ii): the reason the modern realist would not commit to the fundamental reality of spin is because it is—as far as we know—idle vis-à-vis the empirical successes of QM. What tells us that spin is inessential to QM is the fact that we have at least one interpretation of QM—the Bohmian interpretation—where spin is not a fundamental property of electrons (or anything else).[16] Instead, apparent 'spin behaviour' is explained in terms of other properties and relations. Of course, many physicists and philosophers of physics prefer some other interpretation, where spin is 'real'. But their reasons for doing so are not going to be motivating for the cautious selective realist, who is mainly motivated by significant empirical successes, especially predictions. Other interpretations have not enjoyed significant empirical successes that the Bohmian interpretation cannot enjoy.[17] What this means, then, is that we should not talk about spin dropping out of Dirac QM without careful qualification of what we really mean by 'spin'. Nor should we talk of spin causing the fine structure splitting without careful qualification. Nor should we say 'The mathematical formalism of the Dirac equation and group theory require the existence of spin' (Morrison [2004], p. 443), and 'spin shows itself as necessary for conservation of angular momentum' (Morrison [2007], p. 546). Not if we want to stay neutral on interpretations of QM (as the careful realist surely must).

---

[14] There are significant complexities when it comes to the nature of this inference, which needn't be explored here; see (Vickers [2016]).
[15] See, for example, (Morrison [2004], pp. 439ff; Pashby [2012], pp. 443ff).
[16] I should perhaps say that there are variants of the Bohmian interpretation where spin is not a fundamental property (which is enough for my purposes); for discussion, see (Pagonis and Clifton [1995]; Bub [1997], Chapter 6; Daumer *et al.* [1997], especially Section 4).

But then what should the realist be committed to, if not spin? We need to ask what is common to all the different interpretations of QM. Certainly they all include a spin operator, but this in itself doesn't entail that spin is a real property (cf. Daumer *et al.* [1997]). At the same time, the spin operator isn't just mathematics: as the Morrison quote tells us, it has a crucial physical role for the relativistic hydrogen atom, ensuring conservation of angular momentum. Conservation of angular momentum is common to every interpretation, but that doesn't mean that every interpretation demands the reality of spin as a fundamental property of the electron; there are other ways to affect the angular momentum, an underdetermination of the relevant physics. In a semi-classical approach to the Dirac equation the relevant physics concerns a classical 'Thomas' spin precession (Keppeler [2003a], p. 43). In the Sommerfeld case, when one introduces relativity, the angular momentum is affected by a precession not of spin, but of electron orbits. And it should be little surprise (absolutely not a 'miracle') that the effect on angular momentum due to spin in Dirac QM, and the effect on angular momentum due to orbital precession in Sommerfeld's theory, are exactly the same. There is no room for manoeuvre when it comes to the precession and the spin, since neither one is actually introduced, by hand, to the theory. Instead each is simply an inevitable consequence of making the theory relativistic.[18]

Biedenharn is helpful here, putting some technical meat on the bones of my claim that precession is to Sommerfeld's theory what the spin is to Dirac's theory. In his original derivation, Sommerfeld handled the orbital precession by moving to a rotating frame of reference, resulting in equations that take the form of normal 'closed' ellipses. Biedenharn ([1983], p. 14) shows that 'Sommerfeld's transformation to a special coordinate frame (in which the 'rosette motion' is closed [...]) is exactly paralleled by an analogous transformation in the Dirac solution'. This transformation in the Dirac case ([1983], p. 27) acts to make the spin 'disappear' in just the same way that the precession disappears in the Sommerfeld case when one views the hydrogen atom from a rotating frame of reference. And the relationship between the two transformations is made especially clear by the fact that the Dirac transformation turns into Sommerfeld's rotating frame of reference in the classical limit of the quantum theory ([1983], p. 28).

---

[17] There are complications when it comes to making Bohmian mechanics relativistic, which are currently being worked out; see, for example, (Dürr *et al.* [2014]). I will assume in this article that it is not impossible to make Bohmian mechanics relativistic.

[18] Schrödinger tried to make QM relativistic in 1925 simply by introducing the Klein–Gordon Hamiltonian. But to make QM properly relativistic, and such that it is not 'in disagreement with the general principles of quantum mechanics' (Morrison [2004], p. 440), we need to adjust the Schrödinger equation in a more fundamental way, leading to the Dirac equation.

If it is accepted that Sommerfeld's precession plays an exactly analogous role in his theory to the role of spin in the Dirac theory, then this bears heavily on the many suggestions in the literature that Sommerfeld succeeded because his error in omitting the spin is cancelled out by another error he made. But what about the 'killer blow' I introduced back in Section 3? Recall that Keppeler ([2003a], Section 9, [2003b], pp. 105–7) apparently shows that two theoretical features—relevant spin rotation angles and the ½ Maslov term— cancel out (or add to an integer). Further he notes (as does Schiller [1962], p. 1108) that if we only include the ½ Maslov term (leaving out the spin) we end up with the wrong fine structure formula; we need to also introduce the spin to get the fine structure formula back on track. Doesn't this show that they do indeed cancel each other out, and cannot both be described as 'idle wheels' vis-à-vis the fine structure formula?

Not necessarily. Keppeler (like Schiller) is working in a semi-classical framework. This is what makes it possible for him to leave out the spin, and proceed with the derivation to show that one then reaches the wrong fine structure formula. In the full Dirac theory, it is impossible to treat the spin as an independent part of the theoretical framework in this way: it is too intimately integrated in the relevant equations.[19] Keppeler's use of semi-classical theory to draw conclusions about the role of spin is also questionable on the grounds that 'spin' means something different within semi-classical theory: it refers to a classical 'Thomas' spin precession. The semi-classical approach is often thought of as a mathematical technique for finding approximate solutions to quantum problems when an exact solution is unachievable. But really, it is better to think of it as a separate theory—'semi-classical mechanics' (see, for example, Child [2014])—that makes various physical claims that contradict Dirac QM, including claims about 'spin'. Thus, when it comes to 'spin', there is a danger inherent in drawing lessons from the semi-classical framework and applying the conclusions to the full Dirac theory.[20]

---

[19] Recall Section 3, including Footnote 10.
[20] There isn't a similar worry when it comes to comparing the BWS quantum conditions with the semi-classical quantum conditions (cf. the penultimate paragraph of Section 4). That step in my

Thus I maintain that Sommerfeld's success is down to the fact that his theory includes sufficient structural truth. The classical mechanics he employed does leave out something very important, but that missing 'extra structure' is totally idle when it comes to the specific issue of the allowed energy levels. And when it comes to the spin, at a certain level of abstraction Sommerfeld's theory includes exactly what it needs to: in both theories, old and new, there is a certain physical feature that contributes the angular momentum required to ensure conservation of angular momentum. The physics is underdetermined, but that isn't an issue if the realist may be permitted to focus on abstract theoretical 'structure'.

## 6 Structure and Realist Commitment

How was Sommerfeld able to derive predictions of extreme quantitative accuracy with such a radically false physical picture of the hydrogen atom? Isn't it a miracle? Certainly it is surprising, but the above discussion gives us various reasons to steer clear of the dramatic word 'miracle'. For one thing, it is clear that Sommerfeld didn't need to get the physics exactly right, so long as the structure of his theory took on a certain form. But more importantly, Sommerfeld didn't even need to get the structure right, in two different ways. First, we need to make a distinction between 'working' and 'idle' structure (cf. Votsis [2011]). Sommerfeld used the BWS quantum conditions, but these are more specific than they need to be (cf. Saatsi [2005], p. 532). Sommerfeld's derivation is just as successful with the more abstract conditions described by: $\oint pdq = (n + 1/2m)h$, $m = 0$ or $m = 1$ (that is, staying neutral on BWS and semi-classical conditions). Second, Sommerfeld's theory was missing some structure that is a crucial part of Dirac QM. But it turns out that the missing structure is idle vis-à-vis the fine structure formula, so it didn't matter that Sommerfeld missed it out.

The sheer flexibility at the theoretical level vis-à-vis deriving the fine structure formula is surprising. The flexibility extends to the metaphysics, physics, and even the basic mathematical structure. It is definitely not the case that there is a great sensitivity of the final predictions to the starting assumptions. This reduces the 'miraculousness' of Sommerfeld's success quite dramatically. One definitely should not make a comparison with various other examples in physics where there is great sensitivity of final predictions to starting assumptions. For example, if I don't get the position of the sun exactly right, my prediction of the exact duration of the next total solar eclipse in Mexico argument concerns sufficient continuity of relevant structure, and semi-classical mechanics includes the relevant structure of QM given that the term on the RHS of Equation (12) is idle vis-à-vis the result in question.

(268 seconds in Nazas on 8 April 2024) will not be exactly right. Such examples can be misleading, since sometimes there is far less sensitivity of final predictions to starting assumptions. And the less sensitivity there is, the less surprising it is that a false theory could lead to true predictions.

In fact, I have only just scratched the surface on the theoretical flexibility vis-à-vis deriving the hydrogen energy levels. Biedenharn ([1983], pp. 21ff.) shows that non-relativistic QM delivers the Bohr energies whether or not one includes the spin. Including the spin in the non-relativistic quantal treatment only affects the degeneracy (the number of different electron states with the same energy). This is similar to the way Bohr hit upon the correct energies (ignoring the fine structure for a moment) when he only made use of circular orbits: before we go relativistic, all orbits with the same major axis have the same energy (Equation (9)). Thus any model of the hydrogen atom that includes at least one such orbit for each principle quantum number $n$ will deliver the same allowed energies (Equation (4)), and thus the same spectral line frequency predictions.[21]

If we take a Bayesian point of view, this all increases the value for $p(E, \neg T)$. In other words, there is a greater probability than you might originally have thought of getting the correct predictions with the wrong theory. A realist updating her credences based on Sommerfeld's success should therefore make a smaller increase than initial appearances would suggest. However, there is only theoretical flexibility in certain respects: there are features of the underlying structure that absolutely must be left alone if the fine structure formula is to be derived. It follows that the realist needs to make different credence adjustments to different features of the theory, following the successful predictions. This doesn't mean simply separating the 'working' and the 'idle' parts of the theory: it's more complicated than that. For example, the BWS conditions definitely were not simply idle vis-à-vis deriving the fine structure formula, but the realist's credence in them should be modified by the realization that there are other quantum conditions that will do the job just as well. In my view, in the face of Sommerfeld's success, even one's degree of belief in Sommerfeld's electron orbits should increase; it is unrealistic to suggest that somebody living at that time could have made a clean distinction between the 'structure' and the physical electron orbits, labelling the former 'working' and the latter 'idle'. But because the mathematical structure is so fundamental to deriving the predictions, so directly involved, our credence in that structure should increase far more than our credence in the orbits (which are more loosely connected to the final predictions). The realist should never have gotten caught up in a discussion of how to separate the 'working' from the 'idle', as if that were a black-and-white issue. In any case of scientific success,

---

[21] For another example of theoretical flexibility, see (Jammer [1966], pp. 93f).

confirmation will confer upon different theoretical elements to different degrees.

Some philosophers will certainly take issue with this suggested separation (even if not a clean separation) between our epistemic position vis-à-vis the structure and our epistemic position vis-à-vis the orbits. It might seem like a post hoc distinction devised purely to preserve scientific realism. Stanford ([2009], p. 171), in particular, has argued that it would have been 'unintelligible' for scientists of the day to accept the wave nature of light but to deny the existence of the aether. Similarly, I expect that Stanford would say of Sommerfeld's theory that to make a realist commitment to the underlying theoretical structure without also making a commitment to the physical picture involved—complete with precessing elliptical orbits—would have been unintelligible to the scientific community during the period 1916–25. However, I am inclined to think that it would not have been considered absurd by everyone, since some physicists (Peter Debye, for example) were increasingly sceptical of electron orbits. But more importantly, I here offer a prescription, not a description, for the epistemic commitment warranted by a scientific success. Scientists back in 1916 were still closely wedded to classical physics, and hadn't sufficiently left behind the basic assumption that what makes sense at the macroscopic level can also apply at the microscopic level.[22] We have now been accordingly educated, and a purely structural commitment when it comes to fundamental physics makes good sense given that there is nothing remotely intuitive about the quantum world. Further, a focus on 'structure' is already required for a realist response to the multiple different interpretations of QM.

As for the precise details of the structural correspondence between the two theories, I have discussed the close relationship between Hamilton–Jacobi classical mechanics and non-relativistic QM, and also the relationship between the BWS quantum conditions and the semi-classical quantum conditions. Part of the puzzle also concerns the close correspondence between orbital precession in the Sommerfeld theory and spin in the Dirac theory. For the reader looking for further technical details, I can only refer to (Biedenharn [1983]), which provides the physics in all its glory, and concludes that the correspondence between the Sommerfeld derivation and the Dirac derivation is 'the closest possible correspondence' and 'extends to every detail' ([1983], p. 14).[23]

---

[22] Bohr wrote to Carl Oseen in January 1926 that if only the match between theory and experiment had not been so exact, 'then we should not have been tempted to apply mechanics as crudely as we believed possible for some time' (Kragh [1985], p. 85).

[23] Of course, I haven't here provided a positive account of what, exactly, the cautious selective realist faced with Sommerfeld's success would be committed to. But this article is about providing a defence against an historical challenge, where that challenge consists in presenting an apparently radical discontinuity. So to defend against the challenge the realist need only

## 7 Conclusion

I have here presented some central ingredients in a realist defence against the Sommerfeld challenge. In some respects I have provided the essential philosophical discussion required to complement the impenetrable physics and murky dialectic found in (Biedenharn [1983]). For example, in Section 3 I argued that Biedenharn fails to show two things crucial to a realist response: (i) that Sommerfeld's derivation can succeed without reliance upon orbits, and (ii) that spin is not essential to the Dirac derivation (or is hiding in Sommerfeld's derivation). These were tackled in Sections 4 and 5, respectively: (i) was handled by drawing on a structural relationship between classical and QM, and (ii) was handled by noting that (a) Sommerfeld's orbital precession is to classical mechanics what spin is to Dirac QM, (b) spin is not an independent assumption Dirac made—it instead drops out of making QM relativistic, and (c) there is at least one interpretation of QM where spin is not a fundamental property, with the consequence that the cautious realist, who stays neutral on interpretations of QM, must not commit to the reality of spin anyway, regardless of the Sommerfeld puzzle.

Of course both realists and non-realists who despair of 'structural realism' will not be impressed. But if the strongest arguments against my defence are quite general arguments against the viability of structural realism, then that will indicate that the defence is relatively strong. And in addition one needn't be a structural realist tout court in order to formulate a realist defence based on a structural relationship for this particular case (cf. Peters [2014]).

This is a case of great historical and scientific complexity, and no doubt work remains to be done tightening up certain parts of the argument. But if this case is the 'ultimate' historical challenge to scientific realism, then it is already of major significance that the realist can provide a promising answer. Certainly the realist was stunned by this historical challenge. But the comeback is on.

## Acknowledgements

With thanks to audiences at Munich (MCMP, 2016), Durham (2016), UNAM (Mexico, 2016), Nottingham (12th UK workshop on integrated HPS, 2017), Edinburgh (BSPS conference 2017), and Exeter (EPSA conference 2017), where earlier versions of this work were presented. Many thanks in particular to Stefan Keppeler, Juha Saatsi, Margaret Morrison, Lina Jansson, Matthias Egg, Ludwig Fahrbach, Pablo Acuña, Nahuel Sznajderhaus, and two

---

show that there isn't such a radical discontinuity after all, at least in certain crucial respects; see (Vickers [2017]).

**24** *Peter Vickers*

anonymous referees for very helpful discussion and comments. This research was part of the AHRC research project 'Contemporary Scientific Realism and the Challenge from the History of Science', supported by AHRC grant reference AH/L011646/1.

*Department of Philosophy*
*University of Durham*
*Durham, UK*
*peter.vickers@durham.ac.uk*

# References

Arabatzis, T. [2006]: *Representing Electrons*, Chicago, IL: University of Chicago Press.

Biedenharn, L. C. [1983]: 'The "Sommerfeld Puzzle" Revisited and Resolved', *Foundations of Physics*, **13**, pp. 13–34.

Brown, L., Pais, A. and Pippard, A. [1995]: *Twentieth Century Physics*, New York, NY: American Institute of Physics Press.

Bub, J. [1997]: *Interpreting the Quantum World*, Cambridge: Cambridge University Press.

Child, M. S. [2014]: *Semiclassical Mechanics with Molecular Applications*, Oxford: Oxford University Press.

Daumer, M., Dürr, D., Goldstein, S. and Zanghi, N. [1997]: 'Naïve Realism about Operators', *Erkenntnis*, **45**, pp. 379–97.

Duncan, A. and Janssen, M. [2014]: 'The Trouble with Orbits: The Stark Effect in the Old and the New Quantum Theory', *Studies in History and Philosophy of Modern Physics*, **48**, pp. 68–83.

Dürr, D., Goldstein, S., Norsen, T., Struyve, W. and Zanghi, N. [2014]: 'Can Bohmian Mechanics Be Made Relativistic?', *Proceedings of the Royal Society A*, **470**, pp. 20130699.

Eckert, M. [2013]: *Arnold Sommerfeld: Science, Life and Turbulent Times 1868–1951*, New York, NY: Springer.

Eisberg, R. and Resnick, R. [1985]: *Quantum Physics of Atoms, Molecules, Solids, and Particles*, New York, NY: Wiley.

Fahrbach, L. [2011]: 'Theory Change and Degrees of Success', *Philosophy of Science*, **78**, pp. 1283–92.

Ghins, M. [2014]: 'Bohr's Theory of the Hydrogen Atom: A Selective Realist Interpretation', in L. Magnani (ed.), *Model-Based Reasoning in Science and Technology: Theoretical and Cognitive Issues*, Berlin: Springer, pp. 415–29.

Goldstein, H. [1980]: *Classical Mechanics*, Reading, MA: Addison-Wesley.

Granovski, Y. I. [2004]: 'Sommerfeld Formula and Dirac's Theory', *Physics – Uspekhi*, **47**, pp. 523–4.

Griffiths, D. J. [2004]: *Introduction to Quantum Mechanics*, Delhi: Pearson.

Harker, D. [2013]: 'How to Split a Theory: Defending Selective Realism and Convergence without Proximity', *British Journal for the Philosophy of Science*, **64**, pp. 79–106.

Heilbron, J. and Kuhn, T. [1969]: 'The Genesis of the Bohr Atom', *Historical Studies in the Physical Sciences*, **1**, pp. vi-290.

Heisenberg, W. [1968]: 'Ausstrahlung von Sommerfelds Werk in der Gegenwart', *Physikalische Blätter*, **24**, pp. 530–7.

Hettema, H. and Kuipers, T. A. F. [1995]: 'Sommerfeld's Atombau: A Case Study in Potential Truth Approximation', *Poznan Studies in the Philosophy of the Sciences and the Humanities*, **45**, pp. 273–97.

Jammer, M. [1966]: *The Conceptual Development of Quantum Mechanics*, New York, NY: McGraw-Hill.

Keppeler, S. [2003a]: 'Semiclassical Quantisation Rules for the Dirac and Pauli Equations', *Annals of Physics*, **304**, pp. 40–71.

Keppeler, S. [2003b]: *Spinning Particles: Semiclassics and Spectral Statistics*, New York, NY: Springer.

Keppeler, S. [2004]: 'Die "Alte" Quantentheorie, Spin Präzession Und Geometrische Phasen', *Physik Journal*, **3**, pp. 45–9.

Kragh, H. [1985]: 'The Fine Structure of Hydrogen and the Gross Structure of the Physics Community, 1916–26', *Historical Studies in the Physical Sciences*, **15**, pp. 67–125.

Kronig, R. [1960]: 'The Turning Point', in M. Fierz and V. F. Weisskopf (eds), *Theoretical Physics in the Twentieth Century*, New York, NY: Interscience, pp. 5–38.

Letokhov, V. S. and Johansson, S. [2009]: *Astrophysical Lasers*, Oxford: Oxford University Press.

Mehra, J. and Rechenberg, H. [1982]: *The Historical Development of Quantum Theory*, Volume 1, New York, NY: Springer.

Morrison, M. [2004]: 'History and Metaphysics: On the Reality of Spin', in J. Buchwald and A. Warwick (eds), *Histories of the Electron: The Birth of Microphysics*, Cambridge, MA: MIT Press, pp. 425–49.

Morrison, M. [2007]: 'Spin: All Is Not What It Seems', *Studies in History and Philosophy of Modern Physics*, **38**, pp. 529–57.

Müller-Kirsten, H. J. W. [2012]: *Introduction to Quantum Mechanics: Schrödinger Equation and Path Integral*, Singapore: World Scientific.

Norton, J. [2000]: 'How We Know about Electrons', in R. Nola and H. Sankey (eds), *After Popper, Kuhn, and Feyerabend*, Dordrecht: Kluwer, pp. 67–97.

Pagonis, C. and Clifton, R. [1995]: 'Unremarkable Contextualism: Dispositions in the Bohm Theory', *Foundations of Physics*, **25**, pp. 281–96.

Pashby, T. [2012]: 'Dirac's Prediction of the Positron: A Case Study for the Current Scientific Realism Debate', *Perspectives on Science*, **20**, pp. 440–75.

Pauling, L. and Goudsmit, S. [1930]: *The Structure of Line Spectra*, New York, NY: McGraw-Hill.

Peters, D. [2014]: 'What Elements of Successful Scientific Theories Are the Correct Targets for "Selective" Scientific Realism?', *Philosophy of Science*, **81**, pp. 377–97.

Psillos, S. [1999]: *Scientific Realism: How Science Tracks Truth*, London: Routledge.

Rozental, S. [1967]: *Niels Bohr: His Life and Work as Seen by His Friends and Colleagues*, Amsterdam: North Holland.

Saatsi, J. [2005]: 'Reconsidering the Fresnel–Maxwell Case Study', *Studies in History and Philosophy of Science*, **36**, pp. 509–38.

Sakurai, J. J. [1985]: *Modern Quantum Mechanics*, Redwood City, CA: Addison-Wesley.

Schiller, R. [1962]: 'Quasi-Classical Theory of the Nonspinning Electron', *Physical Review*, **125**, pp. 1100–8.

Sommerfeld, A. [1916]: 'Zur Quantentheorie Der Spektrallinien', *Annalen der Physik*, **356**, pp. 1–94, 125–67.

Sommerfeld, A. [2014]: 'On the Theory of the Balmer Series', *The European Physical Journal*, **39**, pp. 157–77.

Stanford, P. K. [2009]: 'Author's Response', *Metascience*, **18**, pp. 379–90.

Vickers, P. [2012]: 'Historical Magic in Old Quantum Theory?', *European Journal for Philosophy of Science*, **2**, pp. 1–19.

Vickers, P. [2016]: 'Towards a Realistic Success-to-Truth Inference for Scientific Realism', *Synthese*, <https://doi.org/10.1007/s11229-016-1150-9>.

Vickers, P. [2017]: 'Understanding the Selective Realist Defence against the PMI', *Synthese*, **194**, pp. 3221–32.

Votsis, I. [2011]: 'Structural Realism: Continuity and Its Limits', in A. Bokulich and P. Bokulich (*eds*), *Scientific Structuralism*, Dordrecht: Springer, pp. 105–17.

Yourgrau, W. and Mandelstam, S. [1968]: *Variational Principles in Dynamics and Quantum Theory*, London: Pitman.